%% file: Measure_factor_v5_14_clean_uncommented.tex
\documentclass[twoside,twocolumn,9pt]{article}
\usepackage{extsizes}
\usepackage[super,sort&compress,comma]{natbib} 
\usepackage[version=3]{mhchem}
\usepackage[left=1.5cm, right=1.5cm, top=1.785cm, bottom=2.0cm]{geometry}
\usepackage{balance}
\usepackage{mathptmx}
\usepackage{amssymb}
\usepackage{sectsty}
\usepackage{graphicx} 
\usepackage{lastpage}
\usepackage[caption=false]{subfig}
\usepackage[format=plain,justification=justified,singlelinecheck=false,font={stretch=1.125,small,sf},labelfont=bf,labelsep=space]{caption}
\usepackage{float}
\usepackage{fancyhdr}
\usepackage{fnpos}
\usepackage[english]{babel}
\addto{\captionsenglish}{%
  
}
\usepackage{array}
\usepackage{droidsans}
\usepackage{charter}
\usepackage[T1]{fontenc}
\usepackage[usenames,dvipsnames]{xcolor}
\usepackage{setspace}
\usepackage[compact]{titlesec}
\usepackage{hyperref}
\usepackage[export]{adjustbox}
\usepackage{epstopdf}

\definecolor{cream}{RGB}{222,217,201}

\newcommand{\bq}{\mathbf{q}}
\newcommand{\D}{\mathcal{D}}
\newcommand{\ttheta}{\tilde{\theta}}
\newcommand{\tih}{\tilde{h}}
\newcommand{\<}{\langle}
\renewcommand{\>}{\rangle}

\newcommand{\N}{\mathcal{N}}
\newcommand{\br}{\mathbf{r}}

\newcommand{\dd}{\mathrm{d}}
\newcommand{\eg}{{\it e.g., }}
\newcommand{\ie}{{\it i.e.: }}
\newcommand{\kb}{k_B}
\newcommand{\etal}{\textit{et al.}~}

\graphicspath{{./images/}}

\allowdisplaybreaks[1]

\begin{document}

\pagestyle{fancy}
\thispagestyle{plain}
\fancypagestyle{plain}{
\renewcommand{\headrulewidth}{0pt}
}

\makeFNbottom
\makeatletter
\renewcommand\LARGE{\@setfontsize\LARGE{15pt}{17}}
\renewcommand\Large{\@setfontsize\Large{12pt}{14}}
\renewcommand\large{\@setfontsize\large{10pt}{12}}
\renewcommand\footnotesize{\@setfontsize\footnotesize{7pt}{10}}
\makeatother

\renewcommand{\thefootnote}{\fnsymbol{footnote}}
\renewcommand\footnoterule{\vspace*{1pt}%
\color{cream}\hrule width 3.5in height 0.4pt \color{black}\vspace*{5pt}} 
\setcounter{secnumdepth}{5}

\makeatletter 
\renewcommand\@biblabel[1]{#1}            
\renewcommand\@makefntext[1]%
{\noindent\makebox[0pt][r]{\@thefnmark\,}#1}
\makeatother 
\renewcommand{\figurename}{\small{Fig.}~}
\sectionfont{\sffamily\Large}
\subsectionfont{\normalsize}
\subsubsectionfont{\bf}
\setstretch{1.125} 
\setlength{\skip\footins}{0.8cm}
\setlength{\footnotesep}{0.25cm}
\setlength{\jot}{10pt}
\titlespacing*{\section}{0pt}{4pt}{4pt}
\titlespacing*{\subsection}{0pt}{15pt}{1pt}

\fancyfoot{}
\fancyfoot[LO,RE]{\vspace{-7.1pt}\includegraphics[height=9pt]{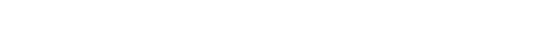}}
\fancyfoot[CO]{\vspace{-7.1pt}\hspace{13.2cm}\includegraphics{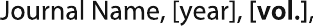}}
\fancyfoot[CE]{\vspace{-7.2pt}\hspace{-14.2cm}\includegraphics{head_foot/RF}}
\fancyfoot[RO]{\footnotesize{\sffamily{1--\pageref{LastPage} ~\textbar  \hspace{2pt}\thepage}}}
\fancyfoot[LE]{\footnotesize{\sffamily{\thepage~\textbar\hspace{3.45cm} 1--\pageref{LastPage}}}}
\fancyhead{}
\renewcommand{\headrulewidth}{0pt} 
\renewcommand{\footrulewidth}{0pt}
\setlength{\arrayrulewidth}{1pt}
\setlength{\columnsep}{6.5mm}
\setlength\bibsep{1pt}

\makeatletter 
\newlength{\figrulesep} 
\setlength{\figrulesep}{0.5\textfloatsep} 

\newcommand{\topfigrule}{\vspace*{-1pt}%
\noindent{\color{cream}\rule[-\figrulesep]{\columnwidth}{1.5pt}} }

\newcommand{\botfigrule}{\vspace*{-2pt}%
\noindent{\color{cream}\rule[\figrulesep]{\columnwidth}{1.5pt}} }

\newcommand{\dblfigrule}{\vspace*{-1pt}%
\noindent{\color{cream}\rule[-\figrulesep]{\textwidth}{1.5pt}} }

\makeatother

\twocolumn[
  \begin{@twocolumnfalse}
{\includegraphics[height=30pt]{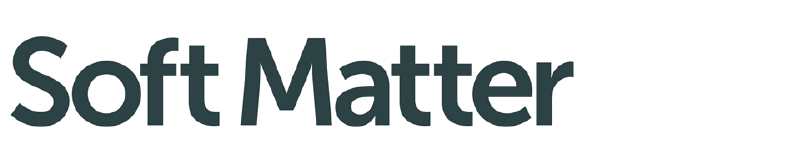}\hfill\raisebox{0pt}[0pt][0pt]{\includegraphics[height=55pt]{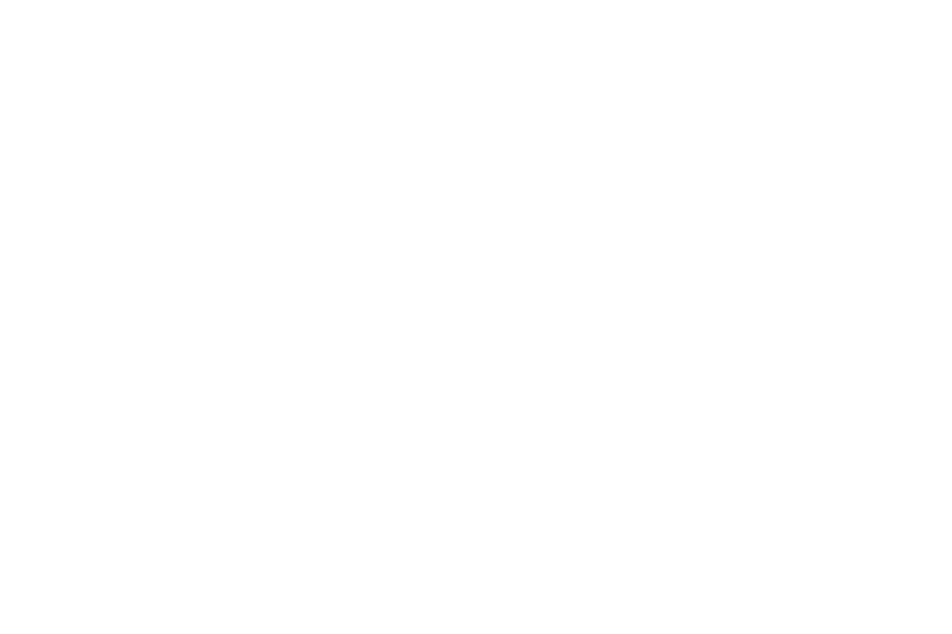}}\\[1ex]
\includegraphics[width=18.5cm]{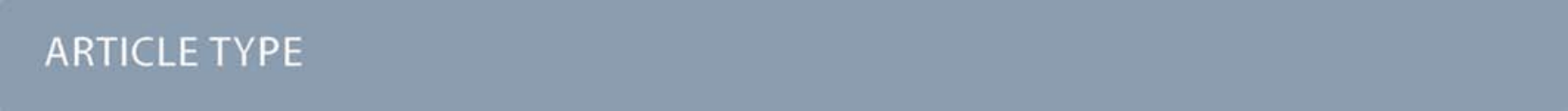}}\par
\vspace{1em}
\sffamily
\begin{tabular}{m{4.5cm} p{13.5cm} }

			\includegraphics{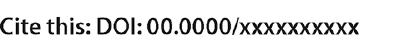} & \noindent\LARGE{\textbf{Frame tension governs the thermal fluctuations of a fluid membrane: a new evidence$^\dag$}} \\
			\vspace{0.3cm} & \vspace{0.3cm} \\
			& \noindent\large{Marc Durand\textit{$^{a}$}} \\
			
			\includegraphics{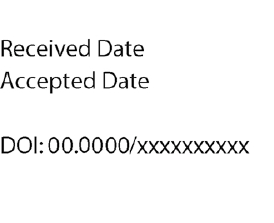} & \noindent\normalsize{Two different tensions can be defined for a fluid membrane: the internal tension, $\gamma$, conjugated to the real membrane area in the Hamiltonian, and the frame tension, $\tau$, conjugated to the projected (or frame) area. According to the standard statistical description of a membrane, the fluctuation spectrum is governed by $\gamma$. However, using rotational invariance arguments, several studies argued that fluctuation spectrum must be governed by the frame tension  $\tau$ instead.
			These studies disagree on the origin of the result obtained with the standard description yet: either a miscounting of configurations, quantified with the integration measure, or the use of a quadratic approximation of the Helfrich Hamiltonian.
			Analyzing the simplest case of a one-dimensional membrane, for which arc length offers a natural parametrization, we give a new proof that the fluctuations are driven by $\tau$, and show that the origin of the issue with the standard description is a miscounting of membrane configurations. The origin itself of this miscounting depends on the thermodynamic ensemble in which calculations are made.
			}	\\		
				\today

		\end{tabular}
		
	\end{@twocolumnfalse} \vspace{0.6cm}
	
	]
	
	\renewcommand*\rmdefault{bch}\normalfont\upshape
	\rmfamily
	\section*{}
	\vspace{-1cm}

	
	\footnotetext{\textit{$^{a}$~Universit\'{e} Paris Cité, CNRS, UMR 7057, Mati\`{e}re et Syst\`{e}mes Complexes (MSC),\\F-75006 Paris, France; E-mail: marc.durand@u-paris.fr}}

	


\section{Introduction}

Bilayer membranes are quasi-two-dimensional
fluid sheets formed by spontaneous self-assembly of lipid
molecules in water. 
They are of the utmost importance for the proper function of living cells, as they spatially separate intracellular compartments and present a boundary to the extracellular environment. Their mechanical properties play a key role in many cellular processes like motility, proliferation and endocytosis.	
	Hence, a physical understanding of the bilayer membrane and its mechanics is necessary for the proper understanding of biological cells, and has
	motivated a long series of investigations. In particular, a number of questions have been raised regarding the definition of the tension of a fluid membrane, stirring up much debate among physicists.

The elasticity of a fluid membrane is traditionally
studied in the framework of the Helfrich effective Hamiltonian \cite{Helfrich_1973, Safran2018}:
\begin{equation}
	\mathcal H= \int_A dA \left\lbrace \gamma + \frac{\kappa}{2} H^2 + \frac{\overline \kappa}{2} G \right\rbrace.
	\label{Helfrich2D}
\end{equation}
Here the integral runs over
the surface area $A$ of the membrane, $H$ and $G$ are the mean and
Gaussian curvatures, while $\kappa$ and $\overline \kappa$ are
the bending rigidity and the Gaussian rigidity, respectively. 
According to Gauss-Bonnet theorem, the surface integral over the Gaussian curvature $G$ can
be written as a constant plus a boundary term which, for larger membranes, is negligible in comparison with the $\gamma A$ term.

It must be noted that Eq. \ref{Helfrich2D} applies to a fluid membrane with fluctuating area $A$: either the membrane is incompressible but exchanges lipid molecules with a reservoir, and then the \textit{intrinsic} (or \emph{internal}) tension
$\gamma$ is essentially the chemical potential of the reservoir of molecules \cite{Cai_1994}; or the membrane is compressible but contains a fixed number of molecules, and then $\gamma$ is the mechanical tension associated with a departure of the lipid surface concentration from the equilibrium configuration \cite{Farago_2011}. 

The intrinsic tension $\gamma$ must not be confused with the \textit{frame} tension $\tau$, which corresponds to the force per unit length that the interface would exert on a surrounding frame whose area $A_p$ is kept fixed, and which is the actual mechanically accessible quantity. Intrinsic and frame tensions are clearly different for a fluctuation area: $\gamma$ is the conjugate variable to the surface area $A$, while $\tau$ is the conjugate variable to its area projected on the frame $A_p=L_p\times L_p$. The latter is then a thermodynamic quantity which depends on the entropy of the membrane, and can be seen as the renormalized version of the former.

When the membrane fluctuates only weakly about a plane, its statistical properties are treated analytically, based on the following set of assumptions:
\begin{enumerate}
	\item Monge parametrization: the membrane position is described in terms of its height $h(x, y)$ above the underlying reference plane as a function of the orthogonal coordinates $x$ and $y$, thereby excluding conformations in which the surface forms overhangs. The area of the membrane is then $A=\iint_{A_p} \sqrt{1+(\nabla h)^2 }dx dy$. 
	\item Assuming small fluctuations, the energy of the membrane is described with the quadratic approximation of the Helfrich Hamiltonian Eq. \ref{Helfrich2D} (omitting
	the contribution of the Gaussian curvature):
	\begin{equation}
		\mathcal H \simeq \iint_{A_p} dx dy \left\lbrace \gamma \left(1+\frac{1}{2}(\nabla h)^2\right)+ \frac{\kappa}{2} (\nabla^2 h)^2  \right\rbrace.
		\label{ApproximatedHelfrich2D}
	\end{equation}
	\item Monge integration measure: the membrane partition function $\Xi$ is obtained by summing over all interface configurations, which mathematically expresses as a functional integral:
	\begin{equation}
		\Xi=\int e^{-\beta \mathcal H[h]}\D[h].
	\end{equation}
	Defining an integration measure $\D[h]$ requires a discretization of the interface \cite{Seifert_concept_1995}. Here, the membrane is discretized in $N_p\times N_p$ patches with constant dimension $a_p$ (with $N_p=L_p/a_p \gg 1$) on the reference frame, and the integration measure is $\D[h]_{\text{Monge}} \equiv \prod_{m,n=1}^{N_p-1} \dd h_{mn}/\ell$ where $\ell$, the \textit{quantum of height fluctuation}, is a characteristic length introduced to render the integration measure properly dimensionless.
\end{enumerate}
In the rest of the present paper, we will refer to this set of assumptions as the \textit{Monge model} (although this is more than just the Monge parametrization, obviously).

A third tension has been introduced in literature, named \emph{fluctuation tension} $r$, and defined as the $q^2-$coefficient in the membrane fluctuation spectrum:
\begin{eqnarray}
	\langle \vert \tilde{h}(\bq) \vert^2 \rangle = \dfrac{k_B T}{A_p \left( r q^2 + \mathcal O ( q^4)\right)}.
	\label{def-r}
\end{eqnarray}
Using very general arguments based on rotational invariance of the free energy of an incompressible membrane, Cai \etal \cite{Cai_1994} came to the conclusion that the fluctuation tension must coincide with the frame tension: $r=\tau$. This result was then extended to the case of a compressible membrane by Farago \cite{Farago_2011}. 
Yet, evaluation of the height fluctuations based on the Monge model yields $r=\gamma$ instead (see Section \ref{non-equiv}).
The reason advanced to explain this discrepancy differs with authors:
Farago  \cite{Farago_2011}, then Schmid \cite{Schmid_2011}, attributes this issue to the use in the Monge model of the approximated Hamiltonian Eq. \ref{ApproximatedHelfrich2D}, which does not satisfy rotational invariance. For Cai \textit{et al.} \cite{Cai_1994} on the other hand, the issue comes from the non-rotational invariance of the integration measure $\D[h]_\text{Monge}$.
Hence, for Farago and Schmid, if one uses the exact Helfrich Hamiltonian but keeps the Monge measure in the calculations, one should recover $r=\tau$, while for Cai \textit{et al.}, the integration measure must be modified. In their study \cite{Cai_1994}, the latter derive the corrective terms to the integration measure in order to recover $r=\tau$,  while still using the approximated Hamiltonian Eq. \ref{ApproximatedHelfrich2D}. 
In spite of the results presented in these different studies, expressions of Eq \ref{def-r} with both  $r=\tau$ \cite{Cai_1994,Borelli_1999,Farago_2004,Schmid_2011,Lindahl_2000,Marrink_2001,Cooke_2005,Wang_2005} and $r=\gamma$\cite{Stecki_2006,Imparato_2006,Fournier_2008,Turlier_2018} are still commonly found in the devoted literature, and still nowadays the exact relation between the different membrane tensions is the object of numerical investigations\cite{Shiba_2016, Gueguen_2017}. 
Elucidating whether the fluctuation tension $r$ matches the frame tension $\tau$, or the intrinsic tension $\gamma$, or none of them, is not just a purely academic question: it is also crucial for experimental \cite{evans_entropy-driven_1990, pecreaux_refined_2004, betz_time_2012} and particle-based numerical \cite{Stecki_2006, Neder_2010, Schmid_2013, chacon_computer_2015, Terzi_2019} investigations dealing with membrane fluctuations. The intrinsic tension, introduced within the Helfrich Hamiltonian, is not a quantity which is directly measurable; its determination relies on accurate theoretical relationship with the frame or fluctuation tension.

In the present paper, focusing on the simpler case of a one-dimensional (1D) membrane embedded in the plane (or equivalently, a two-dimensional membrane which fluctuates along one direction only), we provide a new demonstration that the membrane fluctuations are governed by the frame tension. 
This demonstration enforces previous arguments advanced to justify that $r=\tau$, and then increases our confidence in the final result.
Moreover, our analysis allows to clearly point out the inconsistencies in the Monge model: in agreement with Cai \etal, we show that they are caused by 
an inaccurate integration measure, rather than the use of the approximated Hamiltonian. Nonetheless, restoring rotational invariance of the integration measure is not mandatory to recover $r=\tau$ from calculations: a consistent expansion of the Hamiltonian and the measure leads to the correct result.

The outline of the paper is as follows: in Section \ref{non-equiv}, we give one more evidence than the Monge model has some consistency issues: analyzing the fluctuations of a 1D incompressible membrane either with fluctuating physical length $L$ but fixed projected length on the frame $L_p$, or with fixed $L$ but fluctuating $L_p$, we show that Monge model does not satisfy equivalence of thermodynamic ensembles, in contradiction with fundamentals of statistical physics. Specifically, we obtain $r=\gamma$ in the first ensemble, while $r=\tau$ in the second ensemble. In Section \ref{coarse-graining} we discuss the coarse-graining procedure upon which the Monge model is built, highlighting the origin of the troubles with this model: the choices made for the expression of the coarse-grained Hamiltonian (or its quadratic approximation) and that of the integration measure are not compatible choices to build a free energy that satisfies rotational invariance.
In Section \ref{sec:fixing} we develop a coarse-grained model for a 1D membrane, for which a rotationally invariant integration measure is built from first principles. Using a consistent expansion of this measure an the Hamiltonian, we obtain $r=\tau$ for a membrane with fluctuating length and then recover ensemble equivalence. 
In Section \ref{sec:compress_membrane} we show that to obtain this result, the only correction needed to the Monge model is to take into account the fluctuating character of the number of patches in the integration measure. We then discuss the situation of a compressible membrane with a fixed number of molecules. We show that here again, the Monge integration measure must be modified to recover $r=\tau$. We also discuss the case of an interface between two fluids, which has no bending rigidity and so is stabilized by surface tension only. In that case intrinsic tension and frame tension cannot be distinguised for small fluctuations.

\section{One more evidence of Monge model inconsistency: non-equivalence of ensembles \label{non-equiv}}

We start our analysis by calculating the average square height and average length of a 1D incompressible membrane within two different statistical ensembles using the Monge model, and show that the expressions are different in the thermodynamic limit, in contradiction with fundamentals of statistical physics.

\subsection{$(T,L_p,\gamma)$ ensemble \label{subsectionMonge}}
\begin{figure*}
\centering
\subfloat[]{	
	\includegraphics[width=0.4\textwidth,valign=c]{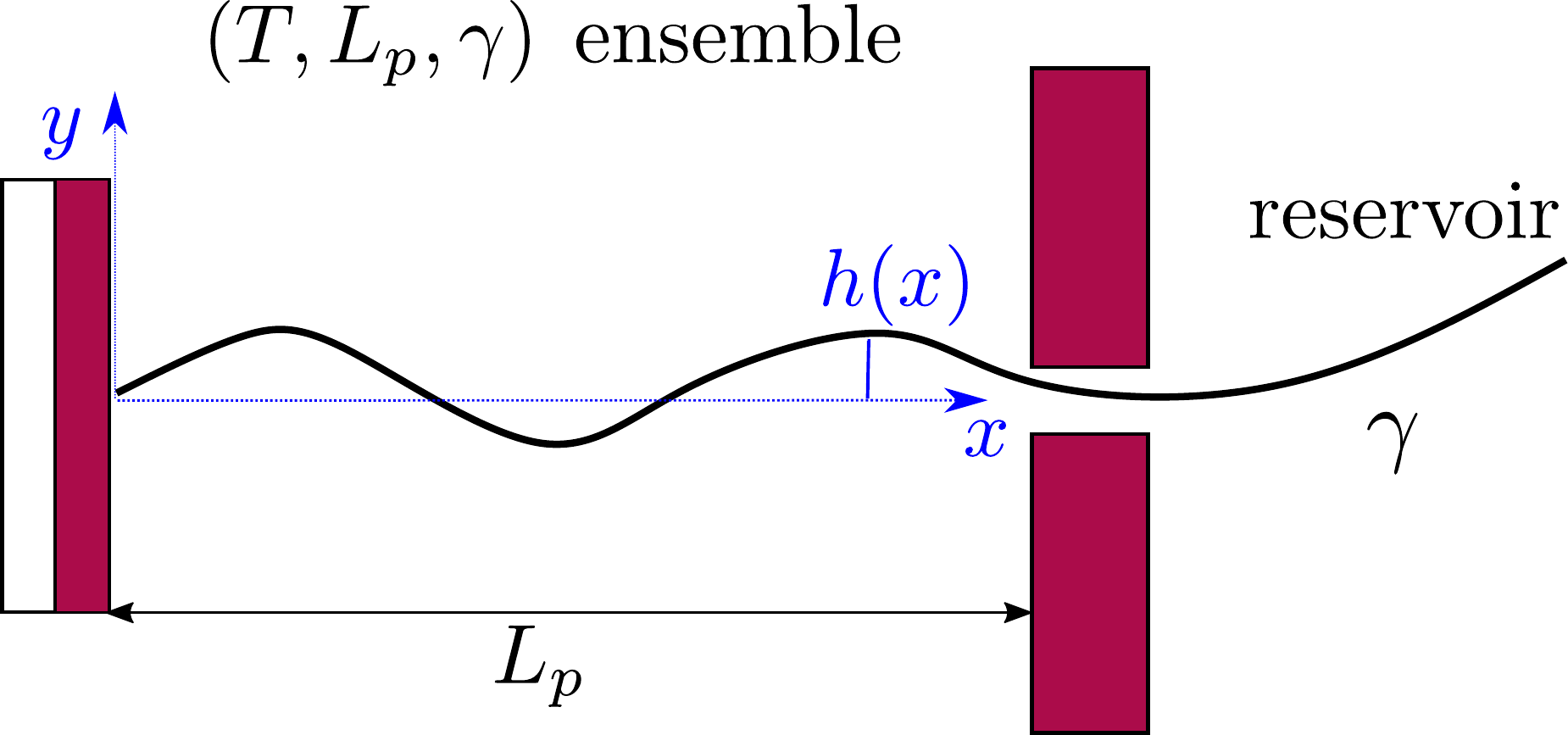}
	\label{fig:T_Lp_Mu}
}
\hfill
\subfloat[]{	
	\includegraphics[width=0.404\textwidth,valign=c]{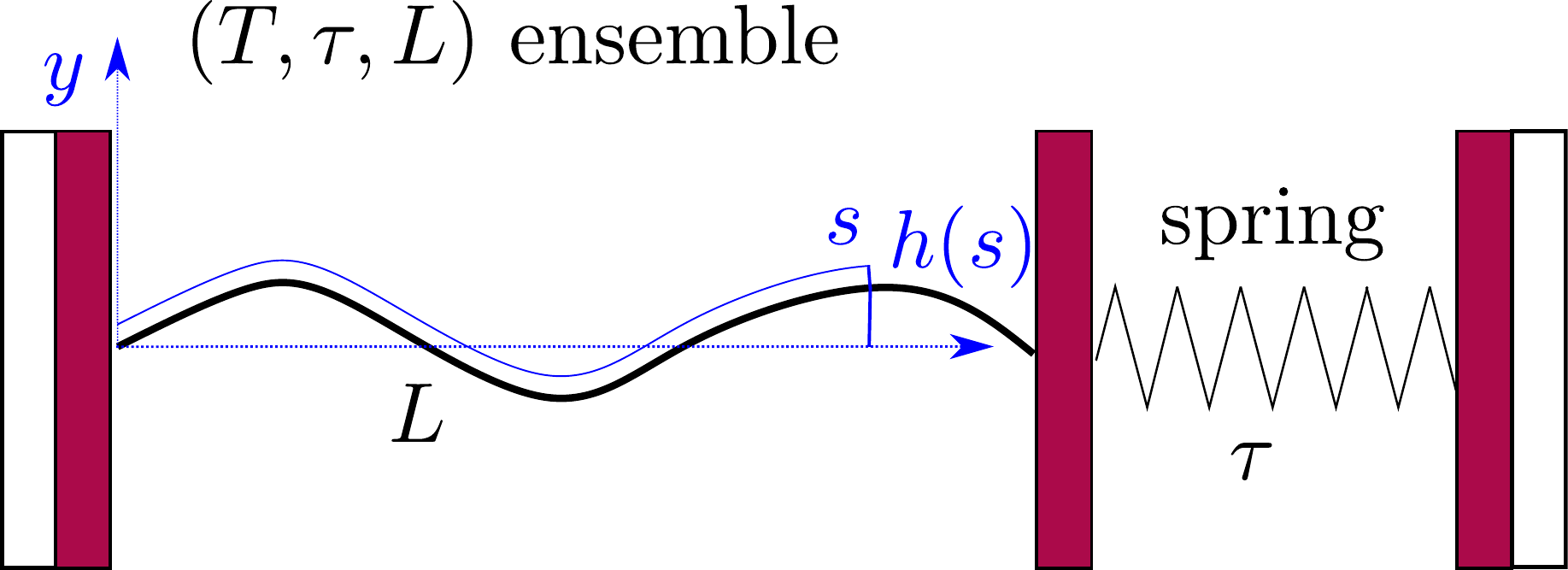}
	\label{fig:T_Tau_L}
}
\caption{Two different thermodynamic ensembles used to calculate the height and length fluctuations of an incompressible membrane: (a) $(T,L_p,\gamma)$ ensemble; (b) $(T,\tau,L)$ ensemble.
}
\end{figure*}

We first consider an incompressible membrane in the ensemble $(T,L_p,\gamma)$ where temperature $T$, projected length $L_p$ and intrinsic tension $\gamma$ are fixed (see Fig. \ref{fig:T_Lp_Mu}). The 1D version of Helfrich Hamiltonian Eq. \ref{Helfrich2D} is
\begin{equation}
	\mathcal H= \int_0^L ds \left\lbrace \gamma + \frac{\kappa}{2} C(s)^2 \right\rbrace,
	\label{Helfrich1D}
\end{equation}
(where $C(s)$ is the local curvature along arc length $s$, and $L$ the length of the membrane), and its quadratic approximation writes
\begin{equation}
	\mathcal H=\gamma L_p +  \int_0^{L_p} dx  \left\lbrace \frac{ \gamma}{2}\left( \dfrac{dh}{dx}\right)^2 + \frac{\kappa}{2} \left(\dfrac{d^2h}{dx^2}\right)^2 \right\rbrace
	\label{Helfrich-approx}
\end{equation}

In agreement with the Monge model, we consider a discretized version of the membrane made of $N_p \gg 1$ patches whose $x$ coordinates are distributed evenly, with projected distance $a_p=L_p/N_p$. Assuming fixed boundary conditions $h(x=0)=h(x=L_p)=0$, we expand $h(x)$ as a sine Fourier series:
\begin{equation}
	h(x)=\sum_{n=1}^{N_p-1} \tih_n \sin \left( \dfrac{n\pi x}{L_p}\right).
	\label{H_reel}
\end{equation}
The Hamiltonian Eq. (\ref{Helfrich-approx}) rewrites in terms of the $\tih_n$ as:
\begin{align}
	\mathcal H=   L_p \left( \gamma + \frac{1}{4} \sum_{n= 1}^{N_p-1} \left(\gamma q_n^2 +\kappa q_n^4 \right) \tih^2_n \right)
	\label{H_Fourier}
\end{align}
where $q_n=n \pi/L_p$.

We then evaluate the partition function of the membrane $\Xi_{\text{Monge}}=\int e^{-\beta \mathcal H }\mathcal D[h]_{\text{Monge}}$, 
with $\mathcal D[h]_{\text{Monge}} \equiv\prod_{n=1}^{N_p-1} \dd h_n /\ell \equiv J_{h \rightarrow \tih} \prod_{n=1}^{N_p-1} \dd \tih_n/\ell$. We introduced in the last equality  
$J_{h \rightarrow \tih}$, the Jacobian associated with the change of variables from the real to reciprocal space. 
Because of the linear relationship between the two sets of variables (Eq. \ref{H_reel}), we know that $J_{h \rightarrow \tih}$ is constant. Its value can be obtained, \eg by calculating a simple Gaussian integral in both direct
and reciprocal spaces. We show in Appendix \ref{Appendix_measure_h} that, with our choice for normalization of Fourier coefficients, one has $J_{h \rightarrow \tih}= ((N_p-1)/2)^{N_p/2}$. 
Summing over all Fourier components $\tih_n$ finally yields
\begin{align}
\Xi_{\text{Monge}} = J_{h \rightarrow \tih}  e^{-\beta \gamma L_p} \prod_{n=1}^{N_p-1} \sqrt{\dfrac{4\pi}{\beta L_p \ell^2(\gamma q_n^2+\kappa q_n^4)}}.
\label{Xi_Monge}
\end{align}
Let us emphasize again that the number of molecules that compose the interface fluctuates, so $\Xi_{\text{Monge}}$ is really a \textit{grand-canonical} partition function.
Assuming moderate tension and/or thermodynamic limit ($\sqrt{\kappa/\gamma} \ll L_p$), and using a continuous description of Fourier modes, it comes
\begin{align}
	&	\ln \Xi_{\text{Monge}} \nonumber \\ & = -\beta L_p \left( \gamma  + \dfrac{k_BT}{2\pi}\int_0^{\pi/a_p} \ln \left[ \dfrac{\beta a_p \ell^2}{2\pi}(\gamma q^2+\kappa q^4) \right] \dd q \right).
	\label{ln_Xi}
\end{align}
Note that the introduction of the Jacobian $J_{h \rightarrow \tih}$ ensures that $\ln \Xi_{\text{Monge}}$ is extensive at the thermodynamic limit: $\ln \Xi_{\text{Monge}} \propto L_p$. It will also affect the expression of the frame tension $\tau$.
We then obtain the average squared amplitude of height straightforwardly: 
\begin{equation}
\< \tih_n^2 \> =  - k_BT \dfrac{\partial \ln \Xi_{\text{Monge}}  }{\partial \alpha_n} = \frac{2k_B T}{L_p(\gamma q_n^2+\kappa q_n^4)},
\label{h2_Monge}
\end{equation}
where $\alpha_n =  L_p (\gamma q_n^2+\kappa q_n^4)/4$. 
Actually, defining the fluctuation tension from the mean square amplitude of Fourier mode is not very convenient when it comes to compare between different thermodynamic ensembles, because the wavenumber does not necessarily follows the same discretization in different ensembles (e.g, multiples of $\pi/L_p$ in ensemble where $L_p$ is fixed, and multiples of $\pi/L$ in ensemble where $L$ is fixed). As a consequence, the variable $\tih_n$ does not correspond to a same wavenumber value in different ensembles. To avoid this complication, we express the average of real variables such as the average squared height $\< {h^2 }\>$ or equivalently, the average membrane length $	\< L \>=L_p\left(1+\<h^{\prime 2}\>/2\right)$ , which can be deduced directly from the partition function as:
\begin{align}
	\< L \> &  = - k_BT \dfrac{\partial \ln \Xi_{\text{Monge}}  }{\partial \gamma} \nonumber \\
	& = L_p\left( 1+\frac{k_BT}{2\pi} \int_0^{\pi/a_p} \dfrac{dq}{\gamma+\kappa q^2} \right)  \label{Lmoy_Monge}\\
	& = L_p\left( 1+\frac{k_BT}{2\pi\sqrt{\kappa \gamma}}\arctan\left(\dfrac{\pi}{a_p}\sqrt{\dfrac{\kappa}{\gamma}}\right) \right).
	\label{Lmoy_Monge_bis}
\end{align}
Eqs \ref{h2_Monge} and \ref{Lmoy_Monge_bis} are the 1D versions of the expressions of $\< \tih_n^2 \>$ and $\< A \>$ derived for a 2D membrane with the Monge model
\cite{Cai_1994, Farago_2003}.
As announced above, within this description the fluctuation tension identifies with the intrinsic tension: $r=\gamma$.
In ref. \cite{Schmid_2011}, expression of $\< L \>$ has been derived assuming $\sqrt{\kappa/\gamma}\gg a_p$, whereas Eq. \ref{Lmoy_Monge_bis} assumes $\sqrt{\kappa/\gamma}\ll L_p$. Hence, both expressions converge to the same following formula within the range $a_p \ll \sqrt{\kappa/\gamma}\ll L_p$:
	\begin{align}
		\< L \>  & \simeq L_p\left( 1+\frac{k_BT}{4\sqrt{\kappa \gamma}}\right).
		\label{Lmoy_Monge_ter}
	\end{align}
The frame tension is given by
\begin{align}
\tau & = -k_B T \left. \dfrac{\partial \ln \Xi_{\text{Monge}}}{\partial L_p} \right|_{T,\gamma}\nonumber \\
& = \gamma  + \dfrac{k_BT}{2\pi}\int_0^{\pi/a_p} \ln \left[ \dfrac{\beta a_p \ell^2}{2\pi}(\gamma q^2+\kappa q^4) \right] \dd q 
\label{tau_vs_gamma}\\
&  = \gamma  + \dfrac{k_BT}{2\pi} \left[ \frac{\pi}{a_p} \ln \left(  \dfrac{\beta a_p \ell^2}{2\pi e^4}\left(\gamma \left( {\pi}/{a_p}\right)^2+\kappa \left( {\pi}/{a_p}\right)^4\right)\right) \right. \\
& \left. +2\sqrt{\frac{\gamma}{\kappa}}  \arctan \left(\frac{\pi}{a_p} \sqrt{\frac{\kappa}{\gamma}} \right) \right].
\label{tau_vs_gamma2}
\end{align}
In the range $a_p \ll \sqrt{\kappa/\gamma}\ll L_p$, verified by most membranes, this expression simplifies to:
	\begin{align}
	\tau & = \gamma \left[1+\dfrac{k_BT}{2\gamma a_p}\ln \left(  \dfrac{\pi^3\kappa  \ell^2}{2 e^4 a_p^3 k_BT}\right)\right].
	\label{tau_vs_gamma3}
	\end{align}


Although the condition $\sqrt{\kappa/\gamma}\gg a_p$ is not required for our purpose, it is easy to see in this regime that $\gamma$ differs from $\tau$: according to Eq. \ref{Lmoy_Monge_ter}, the small fluctuations approximation is satisfied whenever $\kb T/ \sqrt{\kappa \gamma} \ll 1$.  Since $\sqrt{\kappa/\gamma}\gg a_p$, $\kb T/a_p \gamma$ can have any finite value. According to Eq. \ref{tau_vs_gamma3}, one then have $\tau \neq \gamma$ in general (anticipating that $\ell \sim a_p$, as we will argue in Section \ref{sec:fixing}, the argument of the $\ln$ function is $\gg 1$, and so $\tau > \gamma$). The specific case of an interface between two immiscible fluids, for which $\kappa=0$, will be discussed in Section \ref{Interface}.

Note that including quartic terms in the expansion of the Helfrich Hamiltonian would lead to corrective terms into the fluctuation spectrum and thus in the expression of $r$. Using Wick's theorem, it comes that $r=\gamma \left( 1+ \mathcal O \left( k_BT/\gamma L \right) \right)$ \cite{Meunier_1987,Cai_1994}. Clearly, these corrective terms cannot yields $r=\tau$, as this was already emphasized by Cai \textit{et al.}. In the rest of the manuscript we will not consider these corrections coming from the quartic terms in $\mathcal H$.

\subsection{$(T,\tau,L)$ ensemble\label{sec:TtauL}}
The same incompressible membrane is now treated within the $(T,\tau,L)$ ensemble, in which the frame tension $\tau$ and the membrane length $L$ are fixed, while the projected length $L_p$ fluctuates. 
The effective Hamiltonian associated with this ensemble is derived from $\mathcal H$ by a double Legendre transform:
\begin{align}
\mathcal H' & =\mathcal H -\gamma L - \tau L_p
\label{Helfrich2}\\
& =\frac{\kappa}{2} \int_0^L   C^2(s) ds  - \tau \int_0^L  \sqrt{1-\left(\dfrac{dh}{ds}\right)^2} ds
\label{Helfrich-approx2}
\end{align}
and its quadratic approximation writes:
\begin{equation}
\mathcal H'=-\tau L +  \int_{0}^L ds   \left\lbrace \frac{ \tau}{2}\left( \dfrac{dh}{ds}\right)^2 + \frac{\kappa}{2} \left(\dfrac{d^2h}{ds^2}\right)^2 \right\rbrace .
\label{ApproximatedHelfrich1Dbis}
\end{equation}
Note that we have parametrized the height with the arc length $s$ rather than $x$, because $L$ -- and not $L_p$ -- is fixed here. Still using fixed boundary conditions $h(s=0)=h(s=L)=0$, we expand the membrane height as
$h(s)=\sum_{n=1}^{N-1} \tih_n \sin \left(n\pi s/L\right)$. The approximated Hamiltonian then rewrites:
\begin{align}
\mathcal H'=  - L \left( \tau - \frac{1}{4} \sum_{n= 1}^{N-1} \left(\tau q_n^2 +\kappa q_n^4 \right) \tih^2_n \right),
\label{Hprime_Fourier}
\end{align}
where $q_n=n \pi/L$.
After integrating the canonical partition function
$Z=(N/2)^{N/2}\int \exp \left( {-\beta \mathcal H' } \right) \prod_{n=1}^{N-1} \dd \tih_n/\ell$
over the variables $\tih_n$, and still assuming moderate tension or thermodynamic limit ($\sqrt{\kappa/\gamma} \ll L$), we obtain
\begin{equation}
\ln Z = \beta L  \left( \tau - \dfrac{k_BT}{2\pi}\int_0^{\pi/a} \ln \left[ \dfrac{\beta a \ell^2}{2\pi}(\tau q^2+\kappa q^4) \right] \dd q \right),
\label{ln_Z}
\end{equation}
where $a=L/N$ is the patch length. It comes straightforwardly
\begin{equation}
\< \tih_n^2 \> = \frac{2k_B T}{L(\tau q_n^2+\kappa q_n^4)}.
\label{h2_exact}
\end{equation}
Note that this expression is identical to Eq. \ref{h2_Monge} with the substitution $\gamma \to \tau$ and $L_p \to L$. From this observation, ref. \cite{Schmid_2011} (see footnote 5 herein) argued that the fluctuation tension should be renormalized as $r=(L_p/L) \tau$. However, as we already emphasized above, we are comparing the square amplitude of height for two different wavenumbers in the two ensembles, since here $q_n=n\pi/L\neq n\pi/L_p$. Hence, $\< \tih_n^2 \>$ does not correspond to the same quantity in both ensembles. Instead, we compare the relation between membrane and frame length in both ensembles: 
\begin{align}
\< L_p \> &  = k_BT \dfrac{\partial \ln Z  }{\partial \tau}=L\left( 1-\frac{k_BT}{2\pi} \int_0^{\pi/a} \dfrac{dq}{\tau+\kappa q^2} \right)  \\
& = L\left( 1-\frac{k_bT}{2\pi\sqrt{\kappa \tau}}\arctan\left(\dfrac{\pi}{a}\sqrt{\dfrac{\kappa}{\tau}}\right) \right).
\label{Lmoy_exactb}
\end{align}
which can be inverted, in the regime of small fluctuations $(L-L_p)/L_p \ll 1$, as:
\begin{align}
L =	\< L_p \> \left( 1+\frac{k_bT}{2\pi\sqrt{\kappa \tau}}\arctan\left(\dfrac{\pi}{a}\sqrt{\dfrac{\kappa}{\tau}}\right) \right).
\label{Lmoy_exact}
\end{align}
The internal tension $\gamma= k_B T\left. \partial \ln Z/\partial L\right|_{T,\tau}$ expresses here as
\begin{align}
\gamma & =   \tau - \dfrac{k_BT}{2\pi}\int_0^{\pi/a} \ln \left[ \dfrac{\beta a \ell^2}{2\pi}(\tau q^2+\kappa q^4) \right] \dd q. 
\label{gamma_vs_tau}
\end{align}

Eq. \ref{Lmoy_exact} then implies that here $r=\tau$ ! Moreover, following the same arguments as in Section \ref{subsectionMonge}, one can easily realize from Eq. \ref{gamma_vs_tau} that $\gamma\neq \tau$ for finite values of $k_BT/a \tau$.

Note that we took care to distinguish the patch length (named $a_p$ and $a$, respectively) in the two ensembles. It could be argued that the difference between these two cutoffs exactly compensate for the apparent difference in the expression of the average squared height or average length obtained in the two ensembles. A relation between $a$ and $a_p$ can be set by imposing ensemble equivalence for the average energy $\< \mathcal H \>$: according to equipartition theorem, each quadratic term in Eqs. \ref{H_Fourier} and \ref{Hprime_Fourier} contributes $k_BT/2$ to the average energy. Thus, equivalence of ensemble requires that the number of modes in both ensembles, respectively $N_p=L_p/a_p$ and $N=L/a$, are equal. Clearly, with this relation Eqs. \ref{h2_exact} and  \ref{Lmoy_exact} do not match Eqs. \ref{h2_Monge} and \ref{Lmoy_Monge_bis}. Therefore, ensemble equivalence is not satisfied for the average squared height or average length, in the thermodynamic limit ($L, L_p\to \infty$), in contradiction with fundamentals of statistical physics.

Let us emphasize that we assumed $L_p\sqrt{\gamma/\kappa} \gg 1 $ in the calculations, so we are not in the vanishing tension regime, and the equivalence of ensembles is not questionable here \cite{Schmid_2011}.  Indeed, Eq. \ref{Lmoy_exact} (or \ref{Lmoy_Monge_bis}) shows that in this regime, $L$ and $L_p$ are extensive variables which are linearly related for a fixed tension.
In the next two sections we identify the origins of the paradox we pointed out, and show that $r=\tau$ is the correct result indeed.

\section{Tracking the Monge model inconsistencies}\label{coarse-graining}
The description of a membrane at the mesoscopic scale -- where it appears as a continuous sheet -- relies on a coarse-graining construction from its description at the atomic level.  To understand the origin of the issue pointed above, it is useful to come back to the coarse-graining construction at the roots of the Monge model.
As before, we consider a 1D incompressible membrane with fluctuating number of molecules. 
Our starting point is then the classical grand-canonical partition function of the lipid molecules that compose the membrane. After integrating over the molecules' momenta, this writes:
$$\Xi=\sum_{\N}\dfrac{e^{\beta \mu_0 \N }}{\N!}\int\dots\int e^{- \beta U \left( \left\{ \br_\alpha  \right\} \right)}\prod_{\alpha=1}^\N\frac{\dd^2\mathbf{r}_\alpha}{\lambda^2},$$
where $\N$ is the (fluctuating) number of molecules, the $\br_\alpha=(x_\alpha,y_\alpha)$ are their position vectors, $\mu_0$ the chemical potential of the reservoir of molecules, $U\left( \left\{ \br_\alpha  \right\} \right)$ the intermolecular potential, and $\lambda$ the thermal de Broglie length.

We now want to sum over all microstates corresponding to a same membrane shape. 
In a first step, we define a discrete interface by introducing $N_p\ll \N$ points (but $N_p \gg 1$), whose locations
are the coarse-grained \textit{degrees of freedom} (dof) that define the membrane shape.
In the Monge model, these points are distributed evenly on the reference axis $x$, with fixed projected distance $a_p$, and the height $h_{n}$ of point $n$ ($1\leq n \leq N_p$) is defined as the height of the center of mass of the molecules whose $x$ coordinate lies within $ \left[n a_p,(n+1)a_p \right[$. 
Note that this is not the only coarse-grained model we could build, \eg rather than points evenly distributed along the reference frame, we could have chosen them evenly distributed along the membrane geometric support. 
Grouping all microscopic configurations with same heights $\lbrace h_n \rbrace$, the partition function rewrites:
\begin{equation}
\Xi=\idotsint e^{-\beta\mathcal H^\star(\lbrace h_{n} \rbrace)}\prod_{n=1}^{N_p-1}\left(\dd h_{n}/\ell\right),
\label{coarse-grained-Xi}
\end{equation}
where $\mathcal H^\star(\lbrace h_{n} \rbrace)$ is defined through:
\begin{eqnarray}
e^{-\beta\mathcal H^\star(\lbrace h_{n} \rbrace)} =\sum_{\N} & \dfrac{e^{\beta \mu_0 \N }}{\N!} \idotsint \prod_{\alpha=1}^\N  \dfrac {d^{2}\br_{\alpha }}{\lambda ^{2}}  e^{-\beta U \left( \left\lbrace \br_\alpha \right\rbrace \right) } \nonumber \\
&	\times  \prod ^{N_p}_{n=1}   \ell.\delta\left( h_{n}-y_{n}\left( \lbrace \br_{\alpha } \rbrace \right)\right). 
\label{H_star}
\end{eqnarray}
In this equation $y_{n}\left( \lbrace\br_{\alpha } \rbrace \right)$ is the height of the center of mass of the the patch $n$, expressed in terms of the molecules coordinates $ \br_{\alpha }  $:
\begin{eqnarray}
y_{n}\left( \lbrace \br_\alpha \rbrace \right)= \dfrac{ \sum_\alpha \Pi_{n}(x_\alpha)y_\alpha}{\sum_\alpha \Pi_{n}(x_\alpha)},
\end{eqnarray}
where $\Pi_{n}$ is the boxcar function, defined as
\begin{equation}
\Pi_{n}(x)=
\begin{cases}
	1, & \text{if}~ x \in  \left[na_p,(n+1)a_p\right[ \\
	0, & \text{else.}
\end{cases}
\end{equation}
Within the
context of a continuum model, the quantum of height fluctuations $\ell$ introduced in Eq. \ref{coarse-grained-Xi} may be thought of as an arbitrary coarse-graining length for vertical
displacements,
so that the height of a patch center must be increased or decreased by an amount $\ell$ or greater before its new height is treated as distinct from its old one in the sum over membrane configurations  \cite{note_ell}. At the microscopic level, $\ell$ is actually related to the range of the interaction potential $U\left( \left\{ \br_\alpha  \right\} \right)$ between molecules. 

In the continuous limit ($N_p\rightarrow \infty$,  $a_p \rightarrow 0$), $\Xi$ is expressed as a functional integral
\begin{equation}
\Xi=\int e^{-\beta \mathcal H [h]}\mathcal D[h],
\label{Functional-integration}
\end{equation}
where $\mathcal H[h]$ is the Helfrich Hamiltonian (Eq. \ref{Helfrich1D}) and $\mathcal D[h]$ is the measure for functional integration, which should physically correspond to an integration over all interface configurations, not
counting any configuration twice.
Comparing Eqs. \ref{coarse-grained-Xi} and \ref{Functional-integration}, it is tempting to identify $\mathcal H[h]$ with $\mathcal H^\star (\lbrace h_{n} \rbrace)$, and the integration measure $\mathcal D[h]$ with $\D[h]_{\text{Monge}}\equiv\prod_{n=1}^{N_p-1}\left( \dd h_{n} /\ell\right)$. This is actually the identification made in the Monge model. However, other choices are possible: we can introduce \emph{any} function $f(\lbrace h_{n} \rbrace)$ and make the identification $\mathcal D[h] \equiv f(\lbrace h_{n} \rbrace) \prod_{n=1}^{N_p-1}\left( \dd h_{n}/\ell \right)$, $\mathcal H[h] \equiv  \mathcal H^\star (\lbrace h_{n} \rbrace)+k_B T \ln f(\lbrace h_{n} \rbrace)$.
At this point, it may seem artificial or useless to introduce such a function, as the two terms cancel each other. However, the expression of $\mathcal H[h]$ (Eq. \ref{Helfrich1D}) is obtained from an independent reasoning, based on symmetry and invariance properties, and not through the identification with the microscopic description above \cite{Helfrich_1973}. Therefore, the $f$ function appears in the measure only, and it is important to have the correct one to evaluate $\Xi$ correctly. Both $\mathcal H[h]$ and $\mathcal D[h]$ are quantities that should depend solely on the interface shape, and not on the way the surface is parametrized. They should also be invariant under translation and rotation, as $\Xi$. These properties restrain the possible expressions for $f$. Actually, there is something wrong with the choice made within the Monge model  ($f\equiv 1$): the Hamiltonian $\mathcal H^\star$ does not satisfy rotational invariance, because of the Delta Dirac terms in Eq. \ref{H_star}, and thus should not be identified with the (rotationally invariant) Helfrich Hamiltonian $\mathcal H[h]$.
Likewise, the measure  $\D[h]_{\text{Monge}}$ does not satisfy rotational invariance, as pointed out by Cai \textit{et al.} \cite{Cai_1994}.  Hence, there is a clear inconsistency in the Monge model: in one hand, the model use the rotationally invariant Helfrich Hamiltonian for the continuous description of the interface, implying a non-trivial function $f$ to ensure that  $e^{-\beta \mathcal H^\star (\lbrace h_{n} \rbrace)}/ f(\lbrace h_{n} \rbrace)$ is rotationally invariant. On the other hand, the naive integration measure used in the model sets $f(\lbrace h_{n} \rbrace)=1$. Stated differently, identifying $\mathcal H^\star$ with the Helfrich Hamiltonian (or its quadratic approximation) and using  $\D[h]_{\text{Monge}}$
as integration measure are incompatible choices to make $\Xi$ rotationally invariant. It is also clear that replacing the Helfrich Hamiltonian with its quadratic approximation cannot be the (only) cause of the rotational non-invariance of $\Xi$, contrary to what has been argued in previous studies \cite{Farago_2011, Schmid_2011}: the $f$ function that makes $e^{-\beta \mathcal H^\star (\lbrace h_{mn} \rbrace)}/ f(\lbrace h_{mn} \rbrace)$ is still different from $1$.

To restore the rotational invariance of $\Xi$, Cai \textit{et al.} \cite{Cai_1994} introduced two corrective terms to the Monge measure. The first one, named Faddeev-Popov term, corrects the fact that the Monge measure $\D[h]_{\text{Monge}}$  induces 
over-counting of configurations of a fluid membrane. Indeed, the infinitesimal vertical displacement $dh_{n}$ of point $n$ has a component which is locally tangent to the bilayer. At first order, this locally tangent displacement  does not change the shape of the membrane, and then should be disregarded to avoid  multiple counting of a same configuration \cite{note_density_in_Fourier_space}. Faddeev-Popov term regularizes this overcounting by removing the locally tangential component from the measure.
However, Cai \etal acknowledged that this term is still not sufficient to restore the rotational invariance of $\Xi$. They introduced a second term, named Liouville term, to meet this requirement. The interpretation of this term is less obvious, but the above derivation of the coarse-graining procedure helps us to gain insight on its origin: we passed from a fined-grained description where the number of molecules fluctuates, to a coarse-grained description, where the number  $N_p$ of membrane patches is fixed (while their masses fluctuate). But $N_p$ -- or equivalently the projected distance $a_p=L_p/N_p$ -- has been chosen arbitrarily. Its value should be related to the average of the fluctuating number of molecules that compose the interface. An alternative approach, which we will develop in the next section, is to let this number of patches fluctuate by grouping a fixed number of molecules on every patch.

Note that Cai \etal determine the expression of the Liouville term by imposing $r=\tau$, which is required by the rotational invariance of $\Xi$. Hence, the equality $r=\tau$ is not an outcome of the correction, but a requisite instead.
In the next Sections we show that a consistent expansion of the Hamiltonian and the measure -- what requires to take the fluctuating character of the number of patches into account -- allows us to obtain $r=\tau$. 

\begin{figure*}
\centering
\subfloat[]{	
	\includegraphics[width=0.48\textwidth,valign=c]{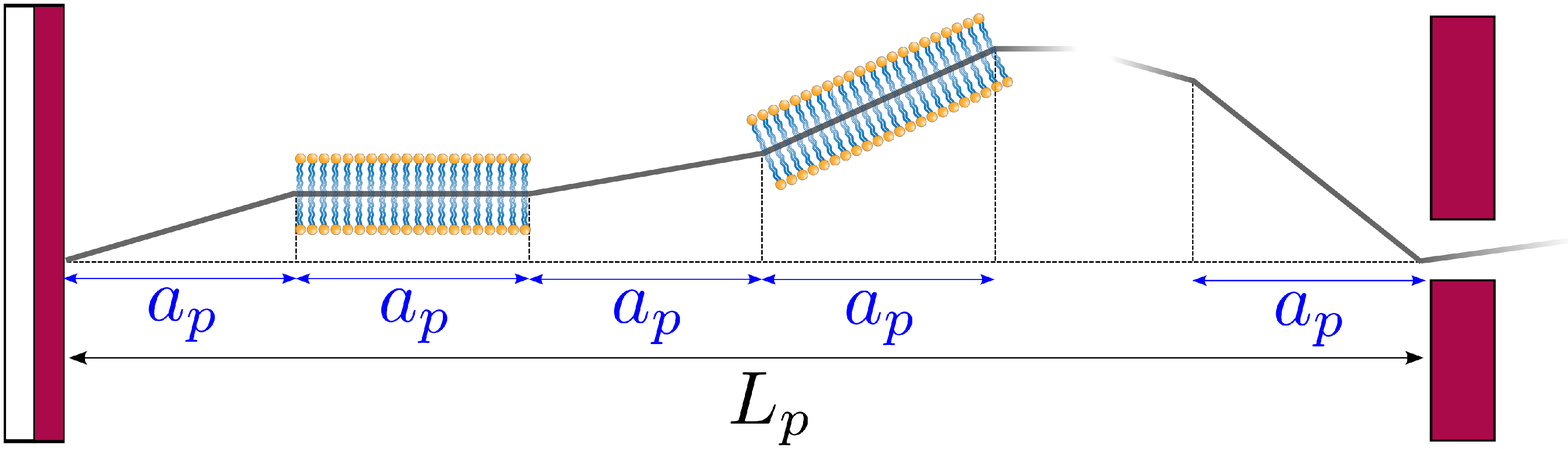}
	\label{fig:bilayer-Monge}
}
\hfill
\subfloat[]{	
	\includegraphics[width=0.48\textwidth,valign=c]{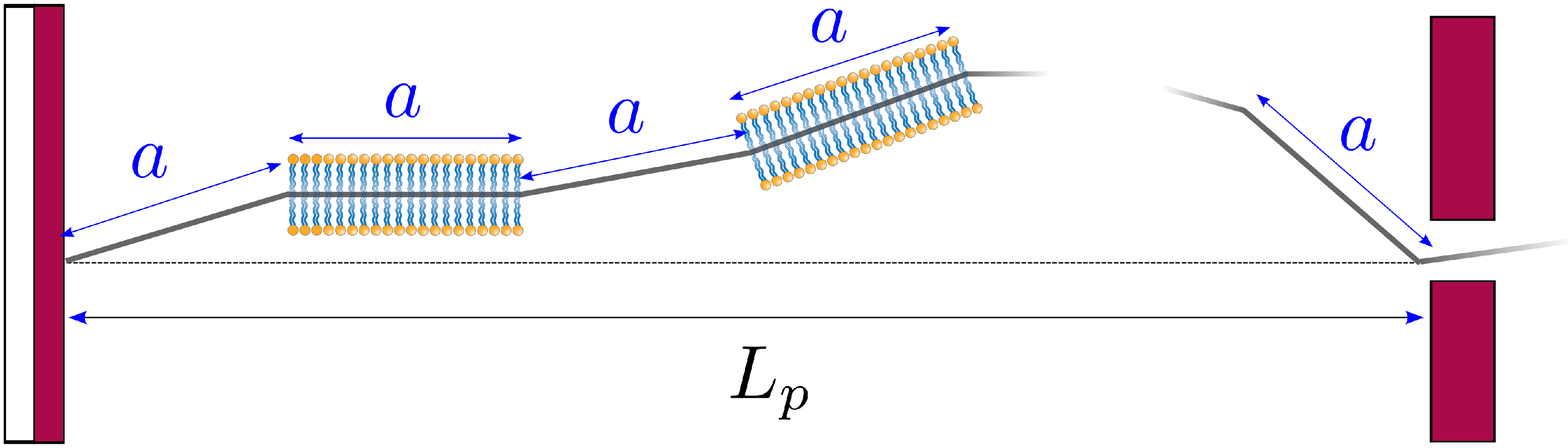}
	\label{fig:bilayer-exact}
}
\caption{Coarse-grained models of an incompressible lipid bilayer in the $(T,L_p,\gamma)$ ensemble. (a) Monge model, in which every patch has a fixed projected length $a_p$, and contains a fluctuating number of molecules; (b) Self-similar coarse-grained description, in which every patch has a fixed length $a$ and contains a fixed number of molecules.
}
\end{figure*}

\section{Fixing Monge model: a self-similar coarse-graining\label{sec:fixing}}

The previous section showed that the Monge model is built on the passage from a fined-grained description in which the number of molecules (with fixed masses) fluctuates to a coarse-grained description in which the number of patches (with fluctuating masses) is fixed. No relation is set on the number of such patches and the average number of molecules in the membrane. 
We develop here an alternate coarse-graining in which each patch contains a fixed number of molecules. The number of patches $N$ is then a fluctuating variable, and considering that the membrane is incompressible, the average number of patches is here trivially related to the average number of molecules in the membrane.
This ``self-similar'' coarse-graining does not alter
the system description: like in the fined-grained description, the membrane is made of a fluctuating number of particles (the patches) with fixed masses, and distributed evenly along the membrane physical length. Furthermore, since patches are all identical, a chemical potential of the patches $\mu$  can be defined unambiguously, and is related to the chemical potential of molecules $\mu_0$ through $\mu=\mu_0 \Gamma a$, where $a$ is the size of a patch and $\Gamma$ the linear density of molecules. 
The definition of a unique chemical potential is less obvious within the Monge model, where patches do not have constant sizes. The intrinsic tension $\gamma$ of the membrane is defined as $\gamma=-\mu /a$.

For a 1D incompressible membrane, patches can only rotate and translate. This is why the 1D case is much simpler than its two-dimensional counterpart, whose patches also deform from one configuration to another, as changes in membrane conformation are generally not isometric. 
	 A coarse-grained configuration of the 1D membrane is then specified by the number of patches it contains $N$, and the orientation $\theta_n$ of each one of them. However, the variables $\theta_n$ cannot vary independently as the projected membrane length of the $x$ and $y$ axes are fixed. Each coarse-grained conformation of the membrane is counted only once in the evaluation of the grand-canonical partition function by summing uniformly over the patch orientations which are compatible with these constraints.
Grouping all fine-grained configurations corresponding to a given value of the variables  $\theta_n$, the grand-canonical partition function becomes:
\begin{equation}
\Xi=  \sum_{N\geq N_{0}}\int e^{-\beta\mathcal H^\star(\lbrace \theta_{n} \rbrace)} \mathcal D_N[\theta],
\label{coarse-grained-Xi-bis}
\end{equation}
where  $N_0=L_p/a$ is the minimal number of patches, and 
\begin{equation}
\mathcal D_N[\theta] \equiv \delta \left(a \sum_{k=0}^{N-1}  \cos \theta_k -L_p\right)\delta \left( a \sum_{k=0}^{N-1}  \sin \theta_k \right)\prod_{n=0}^{N-1}\dd \theta_{n}.
\label{exact-measure}
\end{equation}

In the continuous limit ($N\rightarrow \infty$, $a\rightarrow 0$, with $Na=L$), the membrane is parametrized by the local tangent angle $\theta(s)$ along the curvilinear coordinate $s$ ($s\in [0,L]$).
Clearly, the measure $\mathcal D_N[\theta]$ is invariant by rotation of the reference frame. Subsequently the coarse-grained Hamiltonian $\mathcal H^\star(\lbrace \theta_{n} \rbrace)$ is also rotationally invariant and can be identified with the Helfrich Hamiltonian Eq. \ref{Helfrich1D}, which rewrites in terms of $\theta(s)$ as
\begin{equation}
\mathcal H= \int_0^L ds \left\lbrace \gamma + \frac{\kappa}{2} \left(\frac{d\theta}{ds}\right)^2 \right\rbrace,
\label{Helfrich1D-theta}
\end{equation}

In order to evaluate $\Xi$, we now assume small fluctuations :  $\cos \theta_k \simeq 1-\theta_k^2/2$, $\sin \theta_k \simeq \theta_k$. 
In the continuous limit, the two constraints rewrite as
$$L-\int_0^L  \frac{\theta^2(s)}{2}ds=L_p , \quad \quad \int_0^L  \theta(s)ds=0.$$
Note that this expansion breaks the rotational invariance of the measure $\mathcal D_N[\theta]$, but is consistent with the expression of the Helfrich Hamiltonian Eq. \ref{Helfrich1D-theta}, which is limited to quadratic terms in $\theta(s)$.
Expressing the first constraint with help of the identity
\begin{eqnarray}
\delta(x)=\dfrac{1}{2\pi i}\int_{-i\infty}^{+i\infty}e^{\omega x}d\omega,
\end{eqnarray}
the partition function rewrites:

\begin{eqnarray}
	\Xi \simeq  \int_{L\geq L_{p}} \dfrac{dL}{a} e^{-\beta \gamma L}\int \mathcal{D}_N' [\theta(s)]  e^{-\beta \frac{\kappa}{2}\int_{0}^{L} {\theta'}^2(s) ds} \nonumber \\
	\left[\int_{-i\infty}^{+i\infty}\dfrac{d\omega}{2\pi i} e^{\omega(L-L_p)}  e^{-\frac{\omega}{2} \int_{0}^{L} {\theta}^2(s) ds} \right]
	\label{Z-continu}
\end{eqnarray}
where $\mathcal D_N'[\theta(s)] \equiv \delta \left(\sum_{k}\theta_k\right)\prod_{n=0}^{N-1} d\theta_n$.
Introducing the Fourier expansion $\theta(s)=\sum_{p=1}^{N-1} \ttheta_p \cos \left(p\pi s/L\right)$, in which the zeroth order term has been eliminated to satisfy the second constraint, it comes:
\begin{eqnarray}
	\Xi \simeq  \int_{L\geq L_{p}} \dfrac{dL}{a} e^{-\beta \gamma L} 
	\int_{-i\infty}^{+i\infty}\dfrac{d\omega}{2\pi i} e^{\omega(L-L_p)}	\nonumber \\
	\times	J_{\theta \rightarrow \ttheta} \int  \prod_{p=1}^{N-1} d\ttheta_p \exp \left[-\frac{L}{4}(\beta \kappa q_p^2+\omega)	\sum_{p=1}^{N-1}\ttheta_p^2 \right].
	\label{Z-continu-Fourier}
\end{eqnarray}
We show  in Appendix \ref{Appendix_measure_theta} that the Jacobian associated with the change of variables is still 
$J_{\theta \rightarrow \ttheta} \simeq  (N/2)^{N/2}$ when $N=L/a\gg 1$. Here again, taking the Jacobian into account ensures that $\ln \Xi$ is an extensive quantity for large membranes, and will also affect the expression of the frame tension $\tau$. After integrating over the $\ttheta_p$ variables, we get:
\begin{eqnarray}
\Xi =  \int_{L\geq L_{p}} \dfrac{dL}{a} \int_{-i\infty}^{+i\infty}\dfrac{d\omega}{i2\pi}e^{g(\omega,L)}
\label{Z-continu-2}
\end{eqnarray}
where $g(\omega,L)=-\beta\gamma L+\omega(L-L_p)-L u(\omega)$, with 
$u(\omega)=\int_0^{\pi/a}\ln \left[\frac{a}{2\pi}(\omega +\kappa \beta q^2)\right]dq/2\pi$. Here again, we assumed 
$\sqrt{\kappa/\gamma} \ll L$ so that the lower bound in the integral is $\simeq 0$.

This double integral can be evaluated in the thermodynamic limit ($L, L_p \rightarrow \infty$) thanks to the saddle point approximation:
$\Xi \propto e^{g(\omega^\star,L^\star)}$ where $\omega^\star$  and $L^\star$, the values of $\omega$ and $L$ that extremize $g(\omega,L)$, are given implicitly by:
\begin{equation}
(L^\star-L_p) -L^\star	u'(\omega^\star)=0,
\label{omega_s}
\end{equation}
and
\begin{equation}
\omega^\star=u(\omega^\star)+\beta \gamma.
\label{omega_s2}
\end{equation}
Using this last equation, it comes: $\ln \Xi = g(\omega^\star,L^\star)=-\omega^\star L_p$. 
From Eq. \ref{omega_s2}, we also see that $\omega^\star$ is independent of $L_p$, so the frame tension is
$\tau=-k_BT \partial \ln \Xi/\partial L_p=k_BT \omega^\star$. Injecting this relation into Eq. \ref{omega_s2} finally yields:
\begin{equation}
\gamma= \tau  - \dfrac{k_BT}{2\pi}\int_0^{\pi/a} \ln \left[ \dfrac{\beta a }{2\pi}(\tau  +\kappa q^2) \right] \dd q\\
\label{gamma_vs_tau_2}
\end{equation}
and
\begin{equation}
\ln \Xi = -L_p \beta \left( \gamma +  \dfrac{k_B T}{2\pi}\int_0^{\pi/a} \ln \left[ \dfrac{\beta a }{2\pi}(\tau  +\kappa q^2) \right] \dd q \right).
\label{Xi_exact}
\end{equation}
Eq. \ref{Xi_exact} is identical to Eq. \ref{ln_Xi} derived with the Monge model, just by replacing $\gamma$ with $\tau$ and taking $\ell =e a/\pi$. The quantum of height fluctuations is then of the order of magnitude of a patch size (which is few molecule sizes), while Morse \etal \cite{Morse_1994,Cai_1994} argued it should be of the order of the de Broglie length (which is smaller than molecule size). However, if we had introduced a \emph{quantum of angle fluctuations} $\Theta$ in the integration measure, \ie $\mathcal D_N'[\theta(s)] \equiv \delta \left( \sum_n \theta_n \right)\prod_{n=1}^{N} \left(d\theta_n/\Theta \right)$, we then would end up with $\ell=\Theta e a/\pi$.

Equation \ref{gamma_vs_tau_2} can be inverted for small fluctuations ($k_BT/\sqrt{\kappa \gamma}\ll 1$). The frame tension is thus
\begin{equation}
\tau= \gamma  + \dfrac{k_BT}{2\pi}\int_0^{\pi/a} \ln \left( \dfrac{\beta a }{2\pi}(\gamma  +\kappa q^2) \right) \dd q \left[1+\mathcal O\left(\frac{k_BT}{\sqrt{\kappa \gamma}}\right)\right],\\
\label{tau_vs_gamma2}
\end{equation}
in agreement with previous studies \cite{Cai_1994, Farago_2011,David_vanishing_1991,Borelli_1999}. 

The average length is $\< L \>=-k_BT \partial \ln \Xi/\partial \gamma=k_BT L_p \partial \omega^\star/\partial \gamma$. Deriving Eq. \ref{omega_s2} with respect to $\gamma$ gives: $\partial \omega^\star/\partial \gamma=u'(\omega^\star)\partial \omega^\star/\partial \gamma + \beta$. Together with Eq. \ref{omega_s} this yields $\partial \omega^\star/\partial \gamma=\beta L^\star/L_p$, and thus $\< L \>=L^\star$: the average and most probable values coincide, as expected. Equation \ref{omega_s} then rewrites
\begin{equation}
L_p=\<L\>\left(1- \dfrac{k_B T}{2\pi}\int_0^{\pi/a}\dfrac{dq}{\tau +\kappa  q^2 }  \right).
\label{exact-L_p}
\end{equation}
This equation is identical to Eq. \ref{Lmoy_exactb} derived in the $(T,\tau,L)$ ensemble. Assuming small fluctuations, the relation is inverted as
\begin{equation}
\<L\>=	L_p \left(1+ \dfrac{k_B T}{2\pi}\int_0^{\pi/a}\dfrac{dq}{\tau +\kappa  q^2 }  \right).
	\label{exact-L_p-inverted}
\end{equation}	
	Eq. \ref{exact-L_p-inverted} is then identical to Eq. \ref{Lmoy_Monge} with the substitution $\gamma \to \tau$ (and the undetermined length $a_p$ replaced with the physical patch length $a$).
Using the same substitution in Eq. \ref{h2_Monge}, the height fluctuations are characterized by:
\begin{equation}
\< \tih_n^2 \> = \frac{2k_B T}{L_p(\tau q_n^2+\kappa q_n^4)}.
\end{equation}

These results prove that the fluctuation tension $r$ does coincide with the frame tension $\tau$, in agreement with Cai \etal
It is also worth noticing that Eq. \ref{exact-L_p} suggests a linear scaling of $\<L-L_p\>$ with $\<L\>$ rather than $L_p$ when departing from the regime of small fluctuations, in agreement with the numerical findings of ref. \cite{Schmid_2011}.

\section{Discussion\label{sec:compress_membrane}}
\subsection{Varying patch number is mandatory to recover $r=\tau$} 
Starting from a coarse-grained description in which both the Hamiltonian and the measure are rotationnally invariant, we were able to recover $r=\tau$ for an incompressible membrane within the $(T,L_p,\gamma)$ ensemble.
 However, let us emphasize again that to obtain this result, we used a quadratic expansion of the measure $\mathcal D_N[\theta]$, thereby breaking its rotational invariance. Moreover, the approximation $\sin \theta(s) \simeq \theta(s)$, $\cos \theta(s) \simeq  1-\theta(s)^2/2$ used in the measure is equivalent to assume that $\theta(s)$ and $h(s)$ are linearly related: $\theta(s)\simeq dh/ds$. Therefore, the associated measures are also linearly related: $\mathcal D_N'[\theta] \equiv J_{\theta \rightarrow h}  \ell^N \mathcal D_N[h]$ (the factor $\ell^N$ has been introduced to make the measure dimensionless), with $J_{\theta \rightarrow h}=1/ a^{N} $ (the derivation of this Jacobian is reported in Appendix \ref{Appendix_measure_theta_h}).  Accordingly, the Faddeev-Popov corrective term, which is a correction to the linear relationship between the two measures, and whose expression is derived in Appendix \ref{Appendix_measure_theta_h}, is a second order correction to the expression of $\tau$, in agreement with Cai \etal\cite{Cai_1994}
 Moreover, inserting $\theta(s)\simeq dh/ds$ in Eq. \ref{Helfrich1D-theta} breaks the rotational invariance of the Hamiltonian. Yet, substituting $\theta(s) \rightarrow dh/ds$  (and using $\ell= e a/\pi$) in the grand-canonical partition function (Eq. \ref{Z-continu}) will not change its final expression (Eq. \ref{Xi_exact}). Hence, the use of an approximated Hamiltonian cannot explain the Monge model inaccuracies and the non-equivalence of ensembles discussed in Section \ref{non-equiv} either. This should come as no surprise: ensemble equivalence is a very general feature of standard statistical physics and does not makes any assumption on the realism of the Hamiltonian. It can also be noticed that in the ensemble $(T,\tau,L)$, we obtained  (Sect. \ref{sec:TtauL}) $r=\tau$ while using both an approximated measure and an approximated Hamiltonian.
 
 Therefore, the only remaining reason why the Monge model in the $(T,L_p,\gamma)$ does not give the right value for $r$ is because it disregards the inherent fluctuation of the number of patches. This correction can be identified with the Liouville corrective term introduced in \cite{Cai_1994}. Note that the saddle point approximation we used in our derivation allows us to replace the sum over the patch number with a fixed ad-hoc patch number $N^\star$, and can be interpreted as a correction (so-called Liouville correction) to the Monge measure. However, for calculating accurately other quantities, such as $\<h^4\>$, the other corrections to the Monge model (expansion of $\mathcal H$ to higher terms, and Faddeev-Popov term) might be mandatory.

\subsection{The case of a compressible membrane\label{sec:compressible}}
At this point, one question still remains: we mentioned in the introduction that the same Helfrich Hamiltonian also applies to compressible membranes with fluctuating surface area but fixed number of molecules; the term $\gamma A$ then corresponds to a stretching energy\cite{ note_stretching}.
This is the point of view adopted by Farago and Pincus \cite{Farago_2003, Farago_2011}.  
For a 1D membrane, the relevant thermodynamic ensemble is the ensemble of fixed temperature, number of patches and projected length, $(T,N,L_p)$. How can we then explain that the result obtained with the Monge model ($r=\gamma$) is still not correct in this ensemble ?
Clearly, the issue cannot be associated with a fluctuating number of patches as in the $(T,\gamma,L)$ ensemble. Farago \cite{Farago_2011}, then Schmid \cite{Schmid_2011} invoked the lack of rotational invariance of the approximated Hamiltonian to explain the inaccuracies in the Monge model. Yet, we obtained $r=\tau$ in the $(T,L,\tau)$ and  $(T,L_p,\gamma)$ ensembles while using the very same approximated Hamiltonian, and all three ensembles should be equivalent in the thermodynamic limit.
More likely, the issue originates from the measure $\mathcal D[h]_\text{Monge}$ that still does not properly enumerates the membrane configurations: because the allowed displacements of the patches are those which keep uniform patch length, we should consider all the possible membrane conformations for a given patch length $a$, and sum over all possible values of $a$. The partition function then writes:
\begin{equation}
	\Xi= \int_{a_p}^\infty \dd a \int e^{-\beta\mathcal H[\theta]} \mathcal D_N[\theta],
	\label{Xi-compress}
\end{equation}
where $ \mathcal D_N[\theta]$ and $\mathcal H[\theta] $ are given by Eqs. \ref{exact-measure} and \ref{Helfrich1D-theta}, respectively.
Taking the fluctuating character of the patch size $a$ is thus essential to get the correct result,  the same way that the fluctuation of patch number is essential to recover $r=\tau$ in the  $(T,L_p,\gamma)$. Indeed, in both cases this boils down to sum over membrane length $L$, and the partition function \ref{Xi-compress} is in fact identical to Eq. \ref{Z-continu}. 

\subsection{The case of a fluid interface \label{Interface}}
We finally focus our attention to the case of an interface between two immiscible fluids, and ask ourselves whether internal and frame tensions should be distinguished. At the difference of a membrane, an interface between two immiscible fluids has no bending rigidity ($\kappa=0$) \cite{Safran2018}. The interface is stabilized by the surface tension only, whose origin comes from a non-affinity between molecules of the two bulk phases \cite{Durand2021}, and therefore is very different from -- and usually much larger than -- typical tension value in a membrane.  The high tension criterion $k_B T/\tau a \ll 1$, which is required to be in the small fluctuations regime (Eq. \ref{Lmoy_exact}) when $\kappa \rightarrow 0$, is then satisfied. According to Eq. \ref{tau_vs_gamma2}, $\tau$ and $\gamma$ are then equal up to $\mathcal O ((k_BT/\tau a))^2$ and so can be used indifferently in the expressions of the height and length fluctuations of a fluid interface.

\section{Conclusion}

In summary, starting from the description of a membrane at the molecular level, we first revealed the inconsistencies lying in the foundations of the Monge model to describe the statistics of an incompressible membrane: the choices made for the expression of the coarse-grained Hamiltonian (or its quadratic approximation) and that of the integration measure are not compatible choices to build a free energy that satisfies rotational invariance.
By analyzing the simpler case of a 1D membrane, for which an integration measure that satisfies rotational invariance can be built from first principles, we found in agreement with previous studies that the lack of rotational invariance of the measure used in the Monge model has two origins itself: first, it assumes a fixed number of patches, while for an incompressible membrane with fluctuating length the number of patches must fluctuate; and second, for a given number of membrane patches, there is an over-counting of membrane configurations because the size of a patch is allowed to vary with this linear measure.
Finally, using a consistent expansion of the measure and the Hamiltonian, we were able to prove that the frame tension $\tau$ drives the thermal fluctuations of a membrane. Our analysis then shows that to obtain this result, rotational invariance of the Hamiltonian is not necessary, and one can use its quadratic approximation. Only the measure used in the Monge model must be amended, in agreement with Cai \etal, although its rotational invariance is not required either. In fact, the only correction that must be taken into account is the fluctuating character of the membrane length. 
We believe that the multiplicity of the arguments advanced to justify that $r=\tau$ will reinforce the conviction in these results, and hence will encourage the use of the correct expressions in future studies. 
As a final note, let us emphasize that the Helfrich model represents a simplified version of a real membrane; in particular it is described as an infinitely thin sheet, and energy does not depend on the local lipid concentration along the membrane. All the discussion about whether $r=\tau$ or not is restricted in the context of this specific theoretical framework. But real membranes are more complex, and recent particle-based numerical simulations taking into account the fluctuations of lipid concentration and membrane thickness have been performed in recent years \cite{Neder_2010, Schmid_2013}.  These simulations have shown that for membranes under very strong stretching, the fluctuation tension deviates from the frame tension. A possible explanation for this result is the existence of a coupling term between membrane height and membrane thickness.

\section*{Acknowledgments}
I thank J.B. Fournier for bringing this problem to my attention and for the careful reading of the manuscript.

\section*{Appendix}
\renewcommand{\thesubsection}{\Alph{subsection}}

\subsection{Expression of $J_{h \rightarrow \tih}$\label{Appendix_measure_h}}
The integration measure in the Monge model is $\mathcal D[h]_\text{Monge}\equiv\prod_{n=1}^{N_p-1} \dd h_n /\ell\equiv J_{h \rightarrow \tih} \prod_{n=1}^{N_p-1} \dd \tih_n/\ell$, where $J_{h \rightarrow \tih}$ is the constant Jacobian associated with the linear change of variables Eq. \ref{H_reel}. 
Its value can be obtained e.g., by calculating a simple Gaussian integral $\mathcal I$ in both direct
and reciprocal spaces. Using the identity $\int_0^{L_p}h^2(x)dx=a_p\sum_{n=1}^{N_p-1}h_n^2=(L_p/2)\sum_{n=1}^{N_p-1} \tih_n^2$, the integral writes
\begin{equation}
	\mathcal I=\int \mathcal D[h] e^{-\alpha a_p\sum_n h_n^2} = J_{h \rightarrow \tih} \int \mathcal D[\tih] e^{-\alpha\frac{L_p}{2}\sum_n \tih_n^2},
\end{equation}
where $\alpha$ is any constant with positive real part. The first expression gives $\mathcal I= \left(\pi/\alpha a_p\right)^{(N_p-1)/2}$, and the second one $\mathcal I= J_{h \rightarrow \tih} \left(2\pi/L_p\right)^{(N_p-1)/2}$. Therefore, $J_{h \rightarrow \tih}=(N_p/2)^{(N_p-1)/2}$. 

\subsection{Expression of  $J_{\theta \rightarrow \ttheta}$\label{Appendix_measure_theta}}

We evaluate the Jacobian $J_{\theta \rightarrow \ttheta}$ associated with the linear change of variables $\left\{\theta_n\right\} \rightarrow \left\{\ttheta_p\right\}$, with $\theta_n=\sum_{p=0}^{N-1} \ttheta_p \cos \left(p\pi n/N\right)$, $n \in \{0,...,N-1\}$. As in Appendix \ref{Appendix_measure_h},  $J_{\theta \rightarrow \ttheta}$ is obtained by evaluating a simple integral expressed with both sets of variables. Note however that here the $N$ variables $\theta_n$ are not independent, but are constrained by  $\sum_{n=0}^{N-1}\sin \theta_n \simeq  \sum_{n=0}^{N-1}\theta_n = 0$. This constraint is properly taken into account by introducing a Dirac delta in the integration. For the variables $\lbrace \ttheta_p \rbrace$ on the other hand, this constraint simply reads $\ttheta_0=0$. The second constraint of fixed projected length $a\sum_n \cos \theta_n=L_p$ does not need to be considered to evaluate  $J_{\theta \rightarrow \ttheta}$, since its value does not depend on the choice of the integral to evaluate. Let us then evaluate the integral
\begin{equation}
	\mathcal J = \int e^{- \sum_{n=0}^{N-1}\theta_n^2} \mathcal D'_N[\theta],
\end{equation} 
with $\mathcal D'_N[\theta] \equiv \delta ( \sum_{n=0}^{N-1}\theta_n ) \prod_{n=0}^{N-1}\dd \theta_{n}$. Using the identities $\delta(x)=\int_{-\infty}^{+\infty}e^{iqx} \dd q/2\pi $, and $ \int_{-\infty}^{+\infty} \exp\left[ i q   \theta_n -  \theta_n^2\right]\dd \theta_{n} =e^{-q^2/4}\sqrt{\pi}$, it comes:
\begin{eqnarray}
	\mathcal J & =& \pi^{N/2} \int_{-\infty}^{+\infty} \frac{\dd q}{2\pi }  e^{-Nq^2/4} \\
	&= &\dfrac{ \pi^{(N-1)/2}}{\sqrt{N}} \label{J1}.
\end{eqnarray}
The integral $\mathcal J$ rewrites with variables $\ttheta_p$ as
\begin{eqnarray}
	\mathcal J& =& J_{\theta \rightarrow \ttheta} \int e^{-\frac{N}{2}	\sum_{p=1}^{N-1}\ttheta_p^2} \prod_{p=1}^{N-1}\dd \ttheta_{p} \\ 
	& =& J_{\theta \rightarrow \ttheta} \left(2\pi/N \right)^{\frac{N-1}{2}} \label{J2},
\end{eqnarray}
where we used the equality $\sum_{n=0}^{N-1}\theta_n^2=\dfrac{N}{2}	\sum_{p=1}^{N-1}\ttheta_p^2$, 
Comparing both expressions Eqs. \ref{J1} and \ref{J2} finally yields
\begin{equation}
	J_{\theta \rightarrow \ttheta}=\dfrac{1}{\sqrt 2}\left(\dfrac{N}{2}\right)^{\frac{N}{2}-1}.
\end{equation}
Hence, for $N\gg1$, $\ln J_{\theta \rightarrow \ttheta} \sim \frac{N}{2} \ln \frac{N}{2}$.

\subsection{Expression of $J_{\theta \rightarrow h}$ \label{Appendix_measure_theta_h}}
The two sets of variables are related through $\theta_n=\arcsin \left((h_{n+1}-h_{n})/a\right)$. Since $h_0=h_N=0$, and $\theta_{N-1}=-\sum_{k=0}^{N-2}\theta_k$, there are $N-1$ independent variables for each set. The Jacobian matrix associated with the change of variables $\theta_n \rightarrow h_n$ is
\begin{equation}
	J_{m,n}=\partial \theta_m/\partial h_{n+1}=
	\begin{cases}
		\frac{1}{a\sqrt{1- \left(\frac{h_n-h_{n-1}}{a}\right)^2}} & \text{if }  n=m\\
		\frac{-1}{a\sqrt{1- \left(\frac{h_n-h_{n-1}}{a}\right)^2}}  &\text{if }  n=m-1\\
		0 &\text{else.}
\end{cases}	\end{equation}
and thus its determinant is
\begin{eqnarray}
	J_{\theta \rightarrow h} & = &\dfrac{1}{a^N}\prod_{n=1}^{N-1} \left(1- \left(\frac{h_n-h_{n-1}}{a}\right)^2\right)^{-1/2}\\
	&= &\dfrac{1}{a^N}\exp \left[ -\frac{1}{2}\sum_{n=1}^{N-1} \ln \left(1-  \left(\frac{h_n-h_{n-1}}{a}\right)^2 \right)\right].
\end{eqnarray}
At first order, $\theta(s)\simeq dh/ds$, and then:
\begin{eqnarray}
	J_{\theta \rightarrow h} \simeq 1/a^N.
\end{eqnarray}
Note that expanding the logarithm at next order yields, in the continuous limit:
\begin{eqnarray}
	J_{\theta \rightarrow h} \simeq \dfrac{1}{a^N}\exp \left[ \frac{1}{2a}\int_0^L \left(\frac{dh}{ds}\right)^2 \dd s \right].
\end{eqnarray}
in which we recognize the Faddeev-Popov corrective term \cite{Cai_1994}.


\balance


	\input{Measure_factor_v5_14_clean_uncommented.bbl}

\end{document}

%% file: Measure_factor_v5_14_clean_uncommented.bbl
\providecommand*{\mcitethebibliography}{\thebibliography}
\csname @ifundefined\endcsname{endmcitethebibliography}
{\let\endmcitethebibliography\endthebibliography}{}

%% file: Measure_factor_v5_14_clean_uncommented.bbl
\begin{mcitethebibliography}{33}
\bibitem[not()]{note_ell}
Any free energy contributions arising from vertical displacements of the
  membrane over distances less than $\ell$ are, like the contributions of
  undulation modes of wavelength less than $a_p$, implicitly included within
  the values of the phenomenological parameters of the model, \ie the surface
  tension $\gamma$ and the bending rigidities $\kappa$ and $\bar \kappa$.\relax
\mciteBstWouldAddEndPunctfalse
\mciteSetBstMidEndSepPunct{\mcitedefaultmidpunct}
{}{\mcitedefaultseppunct}\relax
\EndOfBibitem
\bibitem[not()]{note_density_in_Fourier_space}
Hence, the density of interface configurations $h(x,y)$ in the space function,
  whose axes are the $\tih_n$ variables, is not uniform.\relax
\mciteBstWouldAddEndPunctfalse
\mciteSetBstMidEndSepPunct{\mcitedefaultmidpunct}
{}{\mcitedefaultseppunct}\relax
\EndOfBibitem
\bibitem[not()]{note_stretching}
The membrane is under tension ($\gamma > 0$) when the area $A$ is larger than
  the relaxed area $A_0$. We assume that molecules diffuse rapidly along the
  membrane, so that the surface concentration remains uniform. The stretching
  energy then writes for small area departure from $A_0$: $E_s=K(A-A_0)^2/2$,
  where $K$ is the 2D compressive modulus of the membrane. Expanding this
  relation around $A=A_p$, one obtains $E_s\simeq \gamma A +\mathrm{cst}$ with
  $\gamma=K(A_p-A_0)$.\relax
\mciteBstWouldAddEndPunctfalse
\mciteSetBstMidEndSepPunct{\mcitedefaultmidpunct}
{}{\mcitedefaultseppunct}\relax
\EndOfBibitem
\end{mcitethebibliography}
